\documentclass{appolb}
\usepackage{epsfig}

\usepackage{color}
\usepackage{graphicx}
\usepackage{hyperref}

\begin{document}

\newcount\eLiNe\eLiNe=\inputlineno\advance\eLiNe by -1
\title{NA61/SHINE at the CERN SPS:\\ plans, status and first results
}
\author{Antoni Aduszkiewicz\\
for the NA61 collaboration
\address{Faculty of Physics, University of Warsaw,
ul. Ho\.za 69, 00-681 Warszawa, Poland}}
\maketitle

\begin{abstract}
The NA61/SHINE experiment aims to discover the critical point of strongly interacting matter and study properties of the onset of deconfinement. It also to performs precise hadron production measurements for the neutrino and cosmic rays experiments.
These goals will be achieved by measurements of hadron production properties in nucleus-nucleus, proton-proton and pro\-ton-nu\-cle\-us interactions as a function of collision energy and size of the colliding nuclei, as well as pion-nucleus interactions.
This contribution summarises the arguments of the ion program as well as presents the first physics results of the NA61 experiment.
\end{abstract}

\section{Introduction}
\label{introduction}
NA61/SHINE (\emph{SPS Heavy Ion and Neutrino Experiment}) is a fixed-target experiment at the CERN SPS accelerator.
The most important components of the NA61 detector are inherited from the NA49 experiment \cite{Afanasev:1999iu} however several significant upgrades have been introduced (see section~\ref{sec:upgrades}).
The data taking was started in 2007 and is being continued up to now.
The first published and preliminary results are presented in section~\ref{sec_results}.

\section{Physics goals}
\label{Physics goals}
NA61 performs systematic measurements of hadron production in pro\-ton-pro\-ton, proton-nucleus, nucleus-nucleus and pion-nucleus collisions.
The physics program consists of three subjects.
In the first stage of data taking (2007-2010) measurements were performed of hadron production in proton-nucleus interactions needed for neutrino and cosmic ray experiments.
In the second stage (2010-2012) hadron production is studied in proton-proton and proton-nucleus interactions at 158\emph{A}~GeV to obtain reference data for better understanding of nucleus-nucleus reactions.

\begin{figure}
\begin{center}
\includegraphics[width=0.7\columnwidth]{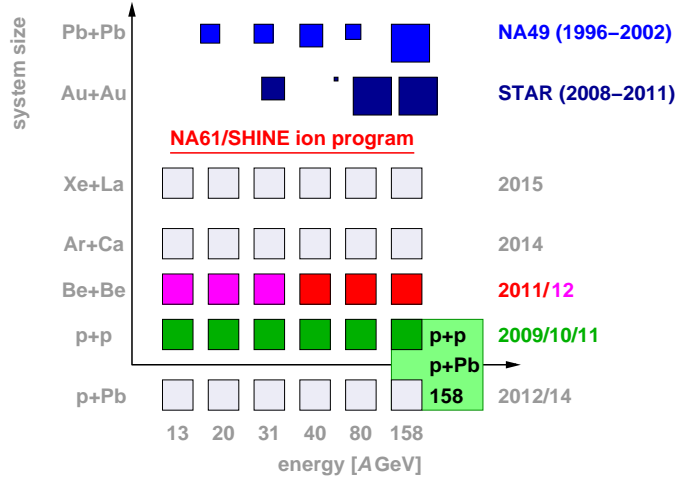}
\end{center}
  \caption{\label{fig:program} Data sets planned to be recorded by NA61/SHINE within the ion program and those recorded by NA49 and STAR.
  The p+p data is already taken. The Be+Be data (40$A$, 80$A$ and 150$A$ GeV) are planned to be taken in November/December 2011.
  The area of the boxes is proportional to the number of registered central collisions, which is typically $2 \times 10^6$ for the SHINE ion program.
}
\end{figure}

In the third stage (2009-2015) the energy dependence of hadron production properties is measured in nucleus-nucleus collisions as well as p+p and p+Pb interactions.
The aim is to identify the properties of the onset of deconfinement and find the critical point of strongly interacting matter.
This requires a comprehensive scan in the whole SPS energy range from 13\emph{A} to 158\emph{A}~GeV with light and intermediate mass nuclei.
NA61/SHINE registers p+p, $^7$Be+$^9$Be, Ar+Ca, Xe+La (the choice of beam and target nuclei is dictated by the technical capabilities of the ion source) and p+Pb collisions at six energies with a typical number of recorded central collision events per reaction and energy of $2 \times 10^6$.
The data planned to be recorded by NA61/SHINE for the ion program and those recorded by NA49 and STAR are compared in Figure~\ref{fig:program}.

\subsection{Study of the properties of the onset of deconfinement}
\label{Study of the properties of the onset of deconfinement}

\begin{figure}
\begin{center}
\includegraphics[width=.49\columnwidth]{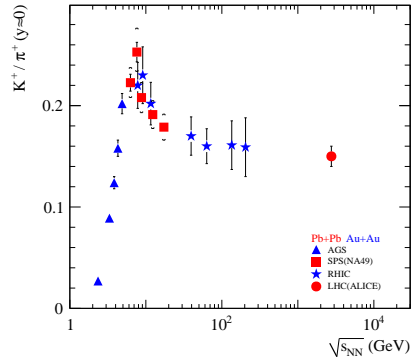}
\end{center}
  \caption{\label{fig:ood} Onset of deconfinement: the ratio of K$^+$  to $\pi^+$ production in A+A collisions change rapidly at the low SPS energies.
}
\end{figure}

\sloppy 

Results on the energy dependence of hadron production in central Pb+Pb collisions at 20$A$-158$A$~GeV coming from the NA49 energy scan program at the CERN SPS provide evidence for the onset of the transition to a system of deconfined quarks and gluons~\cite{afa02,alt08}.
As an example, the energy dependence of the $K^+/\pi^+$ ratio is presented in Fig.~\ref{fig:ood}.
Fig.~\ref{fig:ood} includes also the recent results from the RHIC beam energy scan~\cite{kumar11} as well as the LHC point~\cite{schuk11}.
The RHIC results confirm the NA49 measurements at the onset energies and a smooth evolution is observed between the top SPS (17.2 GeV) and the current LHC (2.76 TeV) energy. For details see \cite{anar11}.

\fussy

The anomalies in hadron production in central Pb+Pb  collisions as a function of collision energy were predicted for the onset of deconfinement~\cite{gaz99} and their further understanding requires new NA61 data.  
The two dimensional scan of NA61/SHINE will provide data between p+p and heavy ion interactions as well as improve the quality of the p+p data.

\subsection{Search for the critical point of strongly interacting matter}
\label{Search for the critical point of strongly interacting matter}
Lattice QCD calculations~\cite{fod04} indicate that the phase diagram of strongly interacting matter features a first order phase transition boundary in the temperature ($T$)-baryochemical potential ($\mu_B$) plane, which has a critical endpoint. This critical endpoint may be located in the energy range accessible at the CERN SPS~\cite{fod04}.

$T$ and $\mu_B$ are not directly measurable quantities but the $T$-$\mu_B$ coordinates of the chemical freeze-out points of nuclear reactions can be brought into one-to-one correspondence with the energy ($E$) and system size ($A$) of the collisions~\cite{bec06}.
Therefore, the $T$-$\mu_B$ coordinates of the chemical freeze-out points may be scanned via a systematic $E$-$A$ scan.
When the freeze-out point is near the critical point, an increase of multiplicity and transverse momentum fluctuations is expected~\cite{ste99}.
The scaled variance $\omega$ of the multiplicity distribution and the $\phi_{p_T}$ measure of transverse momentum fluctuations are expected to increase in the standard NA49 acceptance~\cite{ste99,gre09}.

\section{The NA61 detector}
\label{The NA61 detector}

\begin{figure}
\includegraphics[width=\columnwidth]{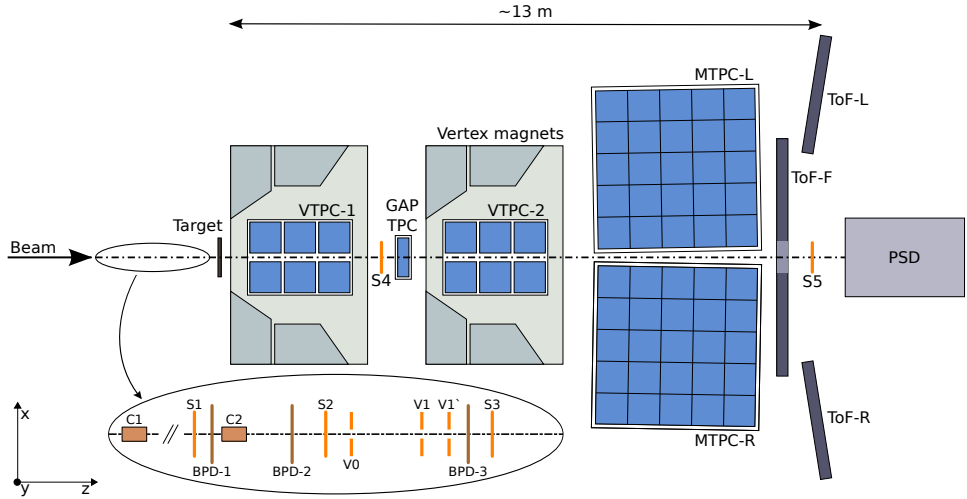}
  \caption{\label{fig:detector} The layout of the NA61/SHINE set-up (top view, not to scale).
  }
\end{figure}

The NA61 experimental setup is shown in Figure~\ref{fig:detector}.
The main components of the NA61 detector are four large-volume Time Projection Chambers for tracking and particle identification.
Two vertex chambers located inside the spectrometer magnets allow to separate positively and negatively charged tracks and to measure the particle momenta precisely.
Two main chambers, placed behind the magnets at both sides of the beam, were optimised for high precision measurement of the energy loss by ionisation.

\subsection{Main upgrades}\label{sec:upgrades}
\label{Main upgrades}

\begin{figure}
\begin{center}
\includegraphics[width=.8\columnwidth]{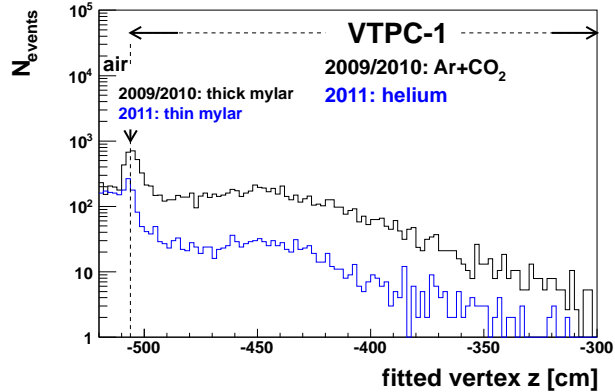}
\end{center}
  \caption{\label{fig:beam_pipe}An illustration of the impact of the new helium-filled beam pipe on the secondary interaction rate.
  The number of secondary vertices is about 10 times lower with the pipe \emph{(blue line)} compared to the old setup \emph{(black line)} when the beam passed through the TPC working gas (Ar/CO$_2$ 90\%/10\%).
  }
\end{figure}

The important NA61/SHINE detector upgrades are motivated by the physics goals.
A new forward time-of-flight detector (ToF-F) was constructed in order to extend the acceptance of the NA61 set-up for pion and kaon identification as required for the T2K measurements.
In order to complete the extensive two dimensional scan program an increase in the event registration rate was necessary.
Speed-up of the ToF-L/R readout, modification of the DAQ system and the new TPC readout enables about 70 Hz readout frequency, which is an increase of a factor of 10.
Furthermore, numerous small modifications and upgrades were performed in order to prepare the detector to work with a broad range of beam types.

One of the upgrades prepared for the ion runs is the construction of the Projectile Spectator Detector (PSD)~\cite{gol09}.
Compared to the NA49 veto calorimeter it provides an increase of the resolution of the measurement of the number of projectile spectators by a factor of 10 to about 
$\sigma(E)/E\approx 50\%/\sqrt E$,
allowing to literally count the spectators. 
The detector consisting of 32 modules is ready for the 2011 Beryllium run. Additional 12 modules will be assembled for the measurements with beams of heavier ions.

Channelling of the high intensity heavy ion beam through the gas volume of the Vertex TPCs has limitations when compared to the proton beam.
Delta electrons from heavy ion beam-gas interactions may significantly increase the background in the TPCs and distort measurements of event-by-event fluctuations.
Therefore a new low-mass helium-filled beam pipe was built and installed in the TPCs before the 2011 run.
It is demonstrated in Fig.~\ref{fig:beam_pipe} that it reduces the interaction background by a factor of about 10.

\subsection{Performance of the detector}
\label{Performance of the detector}
The most important performance parameters of the detector are:
\begin{itemize}
\item Large acceptance $\approx 50\%$ at $p_T <$ 2.5 GeV/c
\item Precise momentum measurement $\sigma (p)/p^2= (0.3 - 7) \times 10^{-4}$ (GeV/c)$^{-1}$
\item High tracking efficiency $>$ 95\%
\item Good particle identification:
  ToF resolution $\sigma(t) \approx 60$ (115) ps in ToF-L/R (ToF-F),
  $dE/dx$ resolution $\sigma (dE/dx) \approx 5\%$,
  invariant mass resolution $\sigma(m) \approx 5$ MeV.
\end{itemize}

\section{Results}\label{sec_results}
\begin{figure}
\begin{center}
  \includegraphics[width=.99\columnwidth]{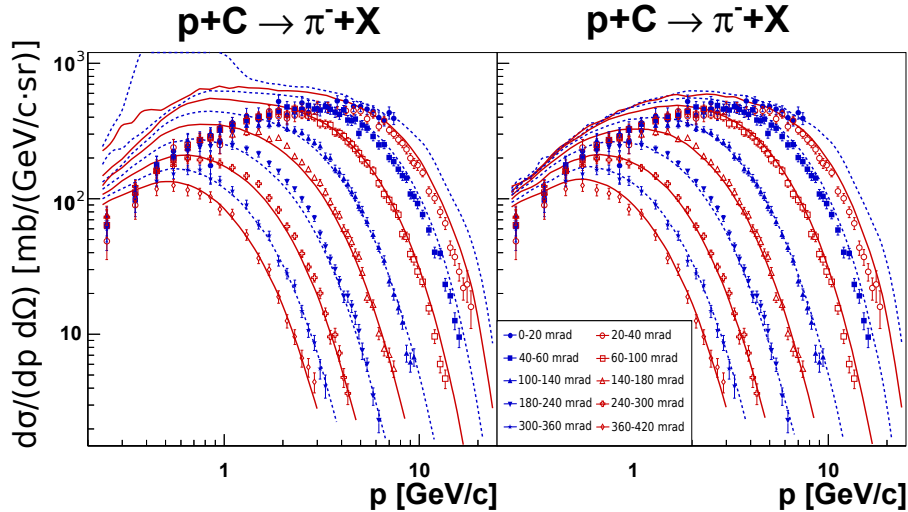}
\end{center}
\caption{\label{fig:urqmd}Inclusive $\pi^-$ spectra measured in NA61/SHINE (points), compared to UrQMD 1.3.1 (lines) \emph{(left)} and patched UrQMD 1.3.1 \emph{(right)}.
}
\end{figure}

The first precise measurements of the charged $\pi$ spectra in p+C interactions at 31 GeV  taken in 2007 were published \cite{na61:pipaper}.
These results have been already used by the T2K neutrino oscillation experiment for the calculations of the neutrino flux \cite{bib:t2k}.
Also, comparison with the UrQMD model revealed discrepancies at low momenta and low polar angles.
The model has been corrected using NA61 results in order to describe the data better \cite{bib:urqmd_patch} as shown in Fig.~\ref{fig:urqmd}.
The measurements of K$^{+}$ spectra from p+C interactions at 31 GeV are to be published soon.

\begin{figure}
\begin{center}
\includegraphics[width=.49\columnwidth]{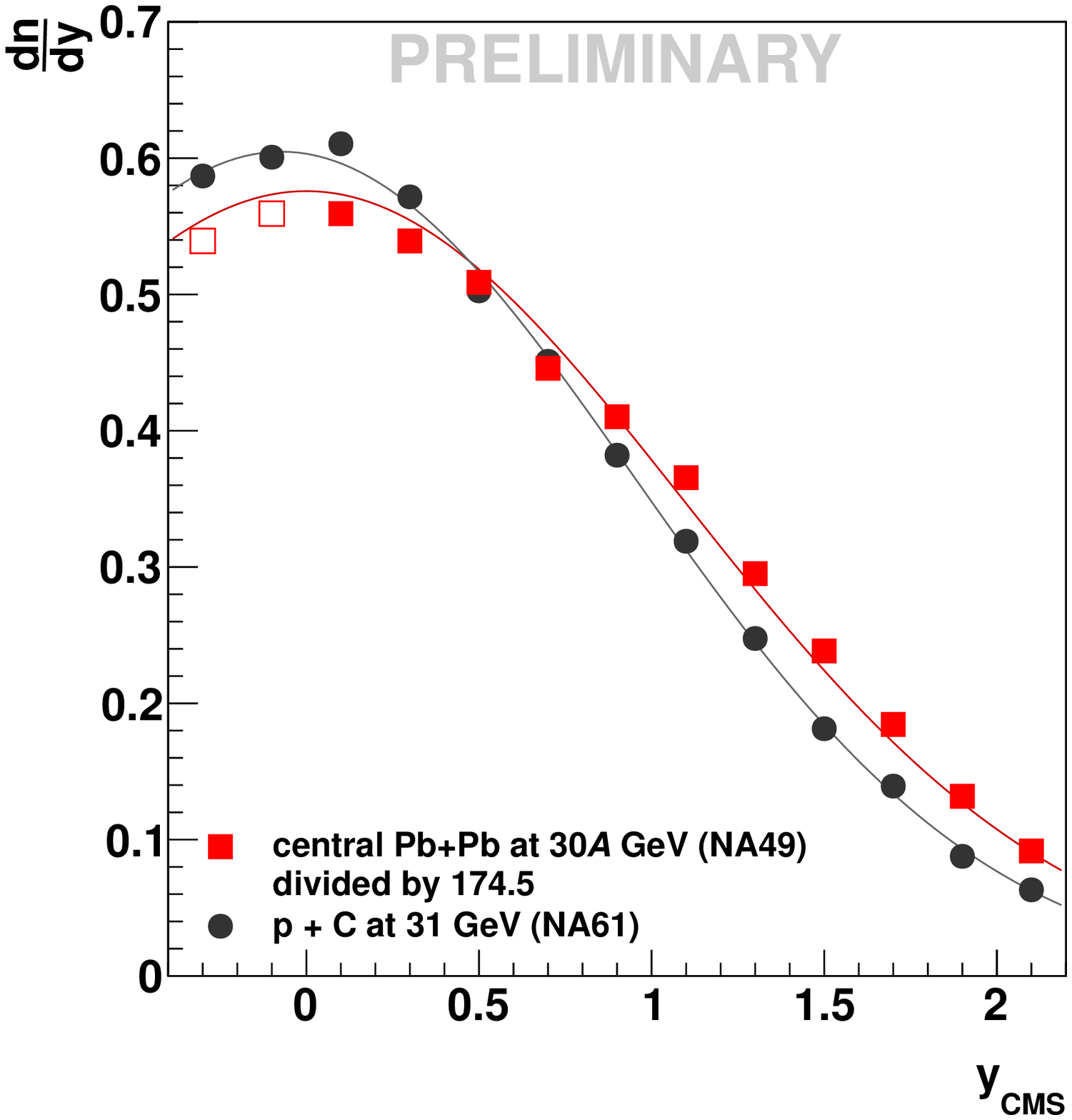}
\includegraphics[width=.49\columnwidth]{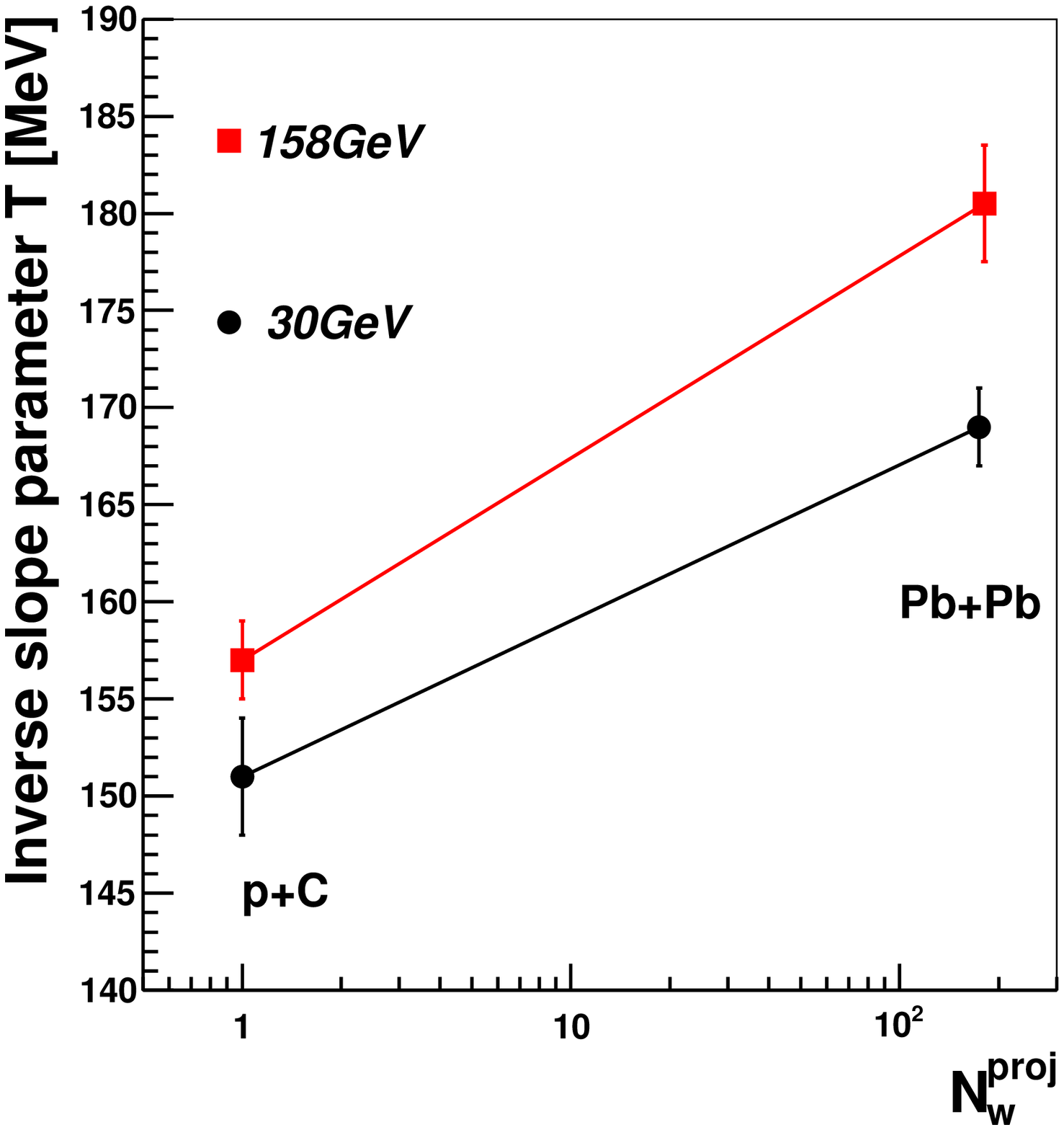}
\end{center}
\caption{\label{fig:results} $\pi^-$ measurements: \emph{(Left):} rapidity distribution in p+C at 31 GeV and central Pb+Pb interactions at 30$A$ GeV for $p_{T}<1$ GeV/c.
The Pb+Pb points are scaled by the predicted wounded projectile nucleon ratio in these two reactions. 
\emph{(Right):} inverse slope parameter of the transverse mass distribution at mid-rapidity for p+C and Pb+Pb at 30$A$/31 and 158($A$) GeV.
}
\end{figure}

First results on $\pi^-$ from NA61 relevant to the study of properties of the onset of deconfinement are presented in Fig. \ref{fig:results}.
The rapidity spectra from p+C interactions at 31 GeV \cite{bib:qm_aduszkiewicz} and central Pb+Pb collisions at 30$A$ GeV \cite{alt08} are compared in Fig. \ref{fig:results} \emph{(left)}.
The p+C rapidity spectrum is shifted towards target rapidity with respect to Pb+Pb collisions due to the projectile-target asymmetry of the initial state.
The mean pion multiplicity in the forward hemisphere is approximately proportional to the mean number of wounded nucleons of the projectile nucleus.
This reflects previous observations~\cite{gaz11} that at the energy of the onset of deconfinement (30A GeV) the mean pion multiplicity in central Pb+Pb collisions agrees with that predicted by the Wounded Nucleon Model~\cite{bial76}.

An exponential function was fitted to the transverse mass spectra at mid-rapidity ($0<y<0.2$) 
\begin{equation}\label{eq_mt}
\frac{1}{m_T}\frac{d^2n}{dm_Tdy} = C\exp\left(-\frac{m_T}{T}\right)
\end{equation}
in the range of $0.2<m_T-m_\pi<0.7$ GeV/c$^2$. The fitted inverse slope parameter $T$ is plotted in Fig. \ref{fig:results} \emph{(right)} for p+C at 31 GeV \cite{bib:qm_aduszkiewicz}, 158 GeV \cite{alt07}, and central Pb+Pb at 30$A$ \cite{alt08} and 158$A$ GeV \cite{alt08andras}.
An increase of this parameter is observed both when changing from proton-nucleus to nucleus-nucleus collisions, and when the collision energy increases.
This can be interpreted as a temperature increase in the statistical models and an increase of radial flow.

\section{Summary}
\label{Summary}
NA61/SHINE has a significant discovery potential for the critical point of strongly interacting matter.
The two dimensional scan of the phase diagram is in progress.
Analysis of the already taken data is ongoing.
p+p interactions were already registered at all 6 energies;
first Be+Be collisions are occuring at the time of writing this text.
NA61 $\pi^\pm$ spectra have been used for computation of the T2K $\nu_\mu$ beam properties, as well as for improving the UrQMD model.

\section{Acknowledgments}
\label{Acknowledgments}
I would like to thank the organizing comitee of the SQM 2011 conference for the invitation.

This work was supported by the Polish Ministry of Science and Higher Education grant N~N202~204638.

Scientific work co-financed by Europejski Fundusz Spo\l{}eczny and the State Budget under ``Zintegrowany Program Operacyjnego Rozwoju Regionalnego'', Dzia\l{}ania 2.6 ``Regionalne Strategie Innowacyjne i transfer wiedzy'' project of Mazowieckie voivodship ``Mazowieckie Stypendium Doktoranckie''.



\end{document}